\documentclass{article}
\usepackage{spconf,amsmath,graphicx}
\usepackage[utf8]{inputenc}
\usepackage{booktabs}
\usepackage{amssymb}
\usepackage{amsfonts}
\usepackage{url}
\usepackage{setspace}
\usepackage{multirow}
\makeatletter
\def\name#1{\gdef\@name{#1\\}}
\makeatother

\title{M3-TTS: Multi-modal DiT Alignment \& Mel-latent for Zero-shot High-fidelity Speech Synthesis}

\name{
\begin{tabular}{c}
Xiaopeng Wang$^{1,\dagger}$,
Chunyu Qiang$^{2,\dagger}$,
Ruibo Fu$^{3,*}$,
Zhengqi Wen$^{3}$,
Xuefei Liu$^{3}$,
Yukun Liu$^{3}$,\\ 
Yuzhe Liang$^{2}$, 
Kang Yin$^{2}$,  
Yuankun Xie$^{3}$,
Heng Xie$^{1}$,
Chenxing Li$^{3}$,
Chen Zhang$^{2}$,
Changsheng Li$^{1}$
\end{tabular}
}
\address{
$^{1}$ Beijing Institute of Technology, Beijing, China 
$^{2}$ Kuaishou Technology, Beijing, China \\
$^{3}$ Institute of Automation, Chinese Academy of Sciences, Beijing, China
}

\begin{document}

\maketitle

\begingroup
  \renewcommand\thefootnote{}
  \footnotetext{$^\dagger$ denotes equal contribution. $^*$ denotes corresponding author.}
\endgroup

\begin{abstract}
Non-autoregressive (NAR) text-to-speech synthesis relies on length alignment between text sequences and audio representations, constraining naturalness and expressiveness. Existing methods depend on duration modeling or pseudo-alignment strategies that severely limit naturalness and computational efficiency. We propose M3-TTS, a concise and efficient NAR TTS paradigm based on multi-modal diffusion transformer (MM-DiT) architecture. M3-TTS employs joint diffusion transformer layers for cross-modal alignment, achieving stable monotonic alignment between variable-length text-speech sequences without pseudo-alignment requirements. Single diffusion transformer layers further enhance acoustic detail modeling. The framework integrates a mel-vae codec that provides 3× training acceleration. Experimental results on Seed-TTS and AISHELL-3 benchmarks demonstrate that M3-TTS achieves state-of-the-art NAR performance with the lowest word error rates (1.36\% English, 1.31\% Chinese) while maintaining competitive naturalness scores. Code and demos will be available at \url{https://wwwwxp.github.io/M3-TTS-Demo}.
\end{abstract}


\section{Introduction}
\label{sec:intro}
In recent years, text-to-speech (TTS) has achieved substantial gains in fidelity and naturalness. Existing methods generally fall into two paradigms: autoregressive (AR) \cite{du2024cosyvoice1,du2024cosyvoice2,maskgct,wang2025sparkttsefficientllmbasedtexttospeech,qiang2024minimally} and non-autoregressive (NAR) \cite{chen2024f5,fastf5,zhu2025zipvoice, E2-TTS, qiang2025instructaudio}. 
AR models generate speech frame by frame \cite {qiang2025secousticodec,qiang2025vq} and thus avoid explicit duration modeling; without hard duration constraints \cite{vall-e,seedtts}, they often produce more expressive prosody and higher naturalness. However, autoregressive decoding results in slow inference, and the teacher-forcing training paradigm induces train–inference mismatch (exposure bias), which can undermine stability. 
By contrast, NAR models typically model the entire acoustic sequence and synthesize speech in parallel, offering significantly faster inference. Nevertheless, most NAR systems depend on text–speech alignment and duration constraints \cite{casanova2022yourtts,maskgct,ju2024naturalspeech,E2-TTS,chen2024f5,zhu2025zipvoice} (e.g., duration predictors, uniform upsampling, or filler padding), which can impose over-regularized timing and pauses and thus average out prosody and expressive variation.

A central challenge for NAR TTS is reliable text--speech alignment. Early solutions employ duration-based aligners: FastSpeech series \cite{ren2019fastspeech,renfastspeech2} depends on forced alignments to provide explicit duration targets, whereas VITS \cite{vits,vits2,casanova2022yourtts} uses monotonic alignment search to infer durations without labeled supervision. Although effective, these strategies impose strong duration constraints that tend to average out prosody and limit expressiveness. Motivated by these limitations, recent Conditional Flow Matching (CFM) \cite{mehta2024matcha} NAR models remove phoneme-level duration modeling. For example, F5-TTS \cite{chen2024f5} and E2-TTS \cite{E2-TTS} pad the text sequence with filler tokens until it matches the mel length, while ZipVoice \cite{zhu2025zipvoice} applies uniform upsampling to assign equal duration to every token. However, both padding and uniform upsampling are surrogate (pseudo-alignment) mechanisms aimed at matching the audio sequence length; such proxy alignment can restrict the model's ability to learn natural timing and rhythm, and it wastes computation by inflating sequences or introducing redundant operations.


In  this paper, we introduce M3-TTS, a non-autoregressive TTS framework that couples a Multi-Modal Diffusion Transformer (MMDiT) architecture with a Mel–VAE latent acoustic target. The model establishes reliable text–speech correspondence without resorting to pseudo-alignment and supports efficient inference; its latent pathway compresses speech in time and dimension, reducing sequence length and memory, stabilizing optimization, and enabling zero-shot synthesis at 44.1 kHz. M3-TTS comprises three elements:

(1) Learnable cross-modal attention performs dynamic, variable-length correspondence between text and speech tokens, eliminating padding or uniform upsampling.

\begin{figure*}[h]
  \centering
  \includegraphics[width=0.9\linewidth]{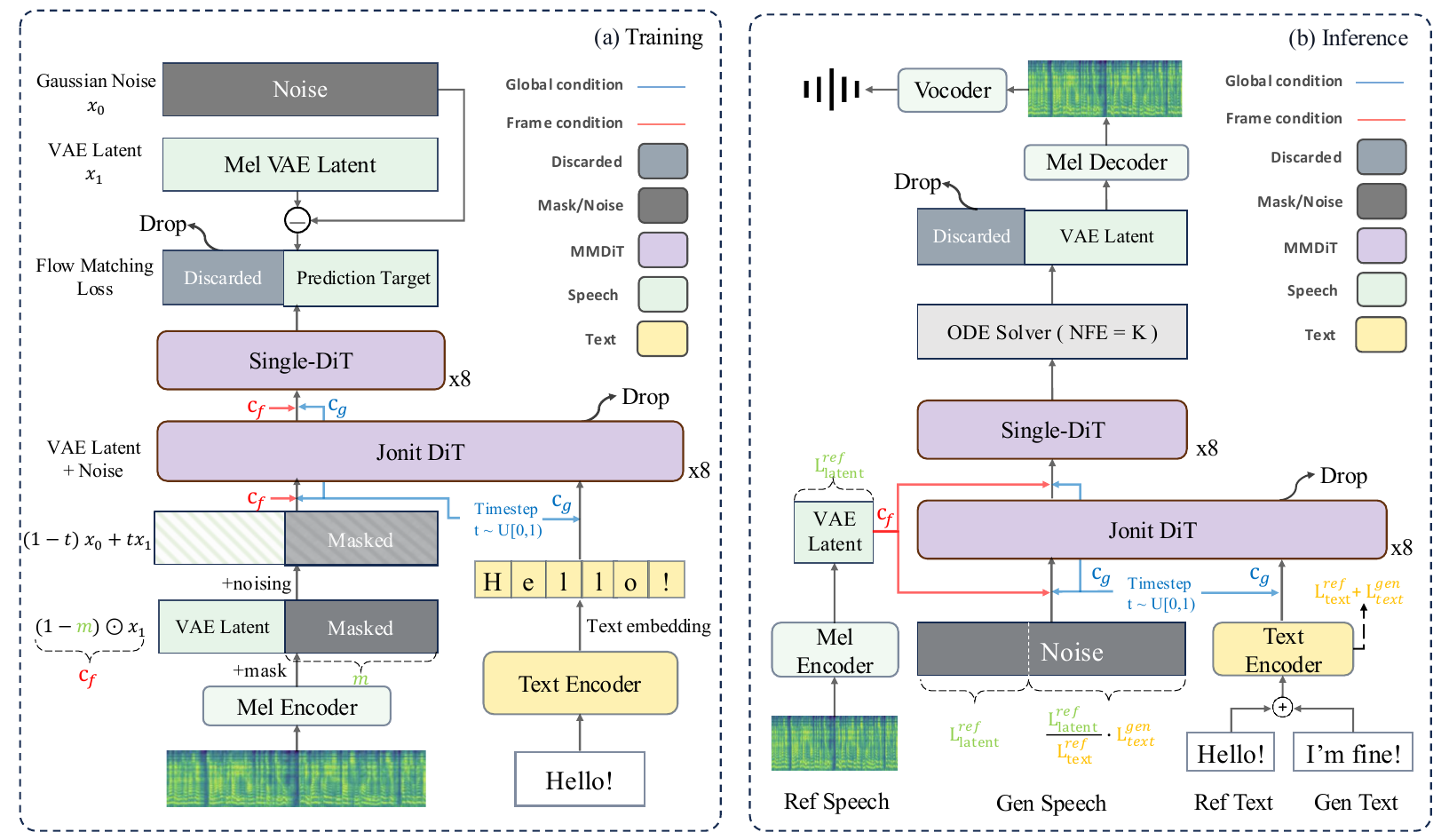}
  \caption{Overview of M3-TTS: training (left) and inference (right).
Text is encoded into $T$, and the reference speech is encoded by a Mel--VAE into latents $x_1$.
Noise $x_0$ and $x_1$ are linearly interpolated to form $x_t$, and conditioning is applied with a global token $c_g$ and a frame-level token $c_f=c_g+(1-m)\odot x_1$.
Joint--DiT aligns $[x_t;T]$ in a unified attention space and splits the output into $(H^{a},H^{t})$.
Single--DiT refines only the speech branch $H^{a}$.
During inference, the output length using the reference speech-to-text ratio; an ODE solver integrates from noise to a latent; the Mel decoder then decodes it into a spectrogram.}
  \label{fig:M3-TTS}
  \vspace{-5pt}
\end{figure*}

(2) The VAE latent provides low-dimensional continuous modeling, shortens both temporal span and feature dimension, lowers GPU memory, improves optimization stability, and natively supports 44.1 kHz synthesis.

(3) Experimental results on Seed-TTS, M3-TTS attains the lowest WER (EN 1.36\%, ZH 1.31\%) among compared systems while maintaining competitive naturalness and speaker similarity.


\section{M3-TTS}
\label{sec:model}

\subsection{Overview}
M3-TTS is designed to overcome the limitations of cross-modal alignment and the inefficiency of high-dimensional mel features. To this end, Mel-VAE compresses speech into a low-dimensional latent space, and a two-stage DiT is adopted: Joint-DiT aligns text representations $T$ with speech latents $x_t$, while Single-DiT refines the speech branch to predict the vector field $v_\theta$ for CFM. During inference, an ODE solver integrates noise into a latent, which is then decoded by the Mel decoder to reconstruct speech. This design avoids padding/upsampling, improves alignment stability and prosody naturalness.

\subsection{Mel--VAE Codec}
The Mel--VAE codec, based on the VQ--CTAP \cite{VQ-CTAP} design and comprising a Mel encoder--decoder pair, normalizes speech to 44.1\,kHz and produces a latent sequence at $\sim$43\, Hz. Compared to mainstream NAR systems, it yields roughly $2\times$ temporal compression and $2.5\times$ dimensional compression ($100\to40$), thereby reducing training/inference memory and compute while maintaining fidelity. Moreover, because the predictor outputs a distribution over latents rather than deterministic log-mel amplitudes, it exhibits improved robustness.

\begin{table*}[!t]
    \centering
    \caption{Objective (SIM-o $\uparrow$, WER $\downarrow$, UTMOS $\uparrow$) and subjective (NMOS $\uparrow$, QMOS $\uparrow$) results on AISHELL3-test (44.1\,kHz) and Seed-TTS test-en/zh (24\,kHz). Compared are AR, NAR, and VAE Reconstruction as the codec upper bound. Bold indicates the best result; underlining denotes the second best.}

    \label{tab:main_results}
    \resizebox{\textwidth}{!}{
    \begin{tabular}{lcc*{3}{c}*{3}{c}*{3}{c}cc}
      \toprule
      \multirow{2}{*}{\textbf{Model}} & \multirow{2}{*}{\textbf{Data (hrs)}} & \multirow{2}{*}{\textbf{Params}} &
      \multicolumn{3}{c}{\textbf{AISHELL3-test (44.1k)}} &
      \multicolumn{3}{c}{\textbf{Seed-TTS test-en (24k)}} &
      \multicolumn{3}{c}{\textbf{Seed-TTS test-zh (24k)}} &
      \multicolumn{2}{c}{\textbf{Subjective Metrics}} \\
      \cmidrule(lr){4-6}\cmidrule(lr){7-9}\cmidrule(lr){10-12}\cmidrule(l){13-14}
      & & & \textbf{SIM-o $\uparrow$} & \textbf{WER $\downarrow$} & \textbf{UTMOS $\uparrow$}
            & \textbf{SIM-o $\uparrow$} & \textbf{WER $\downarrow$} & \textbf{UTMOS $\uparrow$}
            & \textbf{SIM-o $\uparrow$} & \textbf{WER $\downarrow$} & \textbf{UTMOS $\uparrow$}
            & \textbf{NMOS  $\uparrow$} & \textbf{QMOS $\uparrow$} \\
      \midrule
      Ground Truth                    & --            & --   & 0.631 & 6.20  & 2.50 & 0.734 & 2.14 & 3.52 & 0.755 & 1.25 & 2.78 &3.78 ± 0.09 & 3.95 ± 0.08 \\
      VAE Reconstruction              & --            & --   & 0.584 & 5.19 & 1.85 & 0.631 & 1.85 & 2.34 & 0.699 & 1.35 & 1.89 & --   & --   \\
      \midrule
      CosyVoice                       & 170K Multi.   & 416M & --    & --    & --   & 0.609 & 4.29 & --   & 0.723 & 3.63 & --   & --   & --   \\
      CosyVoice 2                     & 167K Multi.   & 618M & --    & --    & --   & 0.652 & 2.57 & --   & 0.748 & 1.45 & --   & --   & --   \\
      Spark-TTS                       & 102K Multi.   & 507M & --    & --    & --   & 0.584 & 1.98 & --   & 0.672 & \textbf{1.20} & --   & --   & --   \\
      \midrule
      E2-TTS (32 NFE)                 & 100K Emilia   & 333M & --    & --    & --   & \textbf{0.706} & 2.32 & 3.21 & 0.713 & 1.91 & 2.26 & --   & --   \\
      F5-TTS (32 NFE)                 & 100K Emilia   & 336M & --    & --    & --   & 0.664 & 1.85 & 3.72 & 0.750 & 1.53 & 2.93 & 3.76 ± 0.20 & 3.90 ± 0.18 \\
      F5-TTS (32 NFE)                 & 100K Emilia   & 155M & --    & --    & --   & 0.628 & 1.96 & 3.66 & 0.733 & 1.57 & 2.93 & --   & --   \\
      ZipVoice (16 NFE)               & 100K Emilia   & 123M & --    & --    & --   & \underline{0.697} & 1.70 & \underline{3.82} & \underline{0.751} & 1.40 & \underline{3.15} & 3.78 ± 0.15 & 3.95 ± 0.13 \\
      \midrule
      M3-TTS-Fbank (32 NFE)           & 100K Emilia   & 355M & --    & --    & --   & 0.681 & \underline{1.48} & \textbf{3.88} & \textbf{0.762} & 1.36 & \textbf{3.18} & \textbf{3.80 ± 0.12} & \textbf{3.99 ± 0.11} \\
      M3-TTS-VAE (32 NFE)             & 100K Emilia   & 355M &  \textbf{0.540}  & \textbf{10.7}  &  \textbf{1.78}   & 0.604 & \textbf{1.36} & 2.80 & 0.621 & \underline{1.31} & 2.18 & 3.62 ± 0.19   & 3.75 ± 0.17   \\
      \bottomrule
      
    \end{tabular}}
  \end{table*}

\subsection{Multi-Modal Diffusion Transformers}
MMDiT architecture comprises two modules: a Joint--DiT for cross-modal alignment and a Single--DiT for speech-only refinement.

\noindent\textbf{Joint--DiT.}
Let speech latents $A\in\mathbb{R}^{B\times T_s\times D}$ and text features $T\in\mathbb{R}^{B\times T_t\times D}$.
We concatenate them along the temporal axis and model the unified sequence with shared attention:
\begin{gather}
Z=\mathrm{Concat}_{\text{time}}(A,T)\in\mathbb{R}^{B\times (T_s+T_t)\times D}.
\end{gather}
Joint--DiT stacks pre-normalized Transformer blocks with \emph{shared} scaled dot-product self-attention and a feed-forward sublayer, aided by modality tags and positional/time embeddings; RoPE is applied to $Q,K$ before attention. Conditions are injected via AdaLN: the text stream uses the global time condition $c_g=\mathrm{Emb}(t)$, while the speech stream uses the frame-level condition:
\begin{gather}
c_f=c_g+(1-m)\odot A,
\end{gather}
where $m\in\{0,1\}^{B\times T_s}$ is a frame-level binary mask (broadcast along the feature dimension).
After cross-modal fusion, the output preserves its shape and is split back into $(H^{a},H^{t})$, yielding explicit text--speech alignment.

\noindent\textbf{Single--DiT.}
We refine only the speech branch $H^{a}$, still conditioned on $c_f$ , and finally output the vector field $v_\theta(\cdot)$ for CFM; the text branch is dropped at this stage.

\subsection{Training and Inference}
\textbf{Training.} We sample $x_0\!\sim\!p_0$ and $t\!\sim\!\mathcal{U}(0,1)$, and form the interpolant $x_t=(1-t)x_0+t\,x_1$. A binary mask $m$ yields a masked view of $x_1$; the global time condition is $c_g=\mathrm{Emb}(t)$ and the fused condition is $c_f=c_g+(1-m)\odot x_1$. With target velocity $u_t=\dot{\alpha}(t)(x_1-x_0)$ (linear schedule $\alpha(t)=t$ reduces to $\dot t(x_1-x_0)$), we minimize:
\begin{gather}
\mathcal{L}(\theta)=\mathbb{E}\Bigl[
\bigl\|\,v_\theta(x_t,t,c_f)-\dot{\alpha}(t)(x_1-x_0)\,\bigr\|_2^2
\Bigr].
\label{eq:fm-loss}
\end{gather}
\textbf{Inference.} The generated latent length is set by:
\begin{gather}
L_{\text{gen}}=\mathrm{round}\!\left(\frac{L^{\text{ref}}_{\text{speech}}}{L^{\text{ref}}_{\text{text}}}\cdot L^{\text{tar}}_{\text{text}}\right).
\label{eq:len}
\end{gather}
Starting from $x_{0}\!\sim\!p_0$, we integrate $\dot{x}_t=v_\theta(x_t,t,c_f)$ for $t\!\in\![0,1]$ to obtain $x_{1}$, then decode it to Mel and waveform.

\section{Experimental Setup}
\label{sec:Experimental Setup}


\noindent\textbf{Training configuration.}
We use a 16-layer MMDiT acoustic model—8 Joint-DiT layers and 8 Single-DiT layers—with model dimension 640 and 10 attention heads ($\approx$355M parameters). The text encoder follows the ZipVoice~\cite{zhu2025zipvoice} design with a 4-layer Zipformer~\cite{yao2023zipformer}. Training is performed on $8\times$A100 GPUs with batch size 192 and learning rate $7.5\!\times\!10^{-5}$. For the infilling objective, we randomly mask 70–100\% of Mel frames. For CFG \cite{cfg} training, masked speech and text inputs are independently dropped with probability 0.2. To verify the advantages of the M3--TTS architecture and keep comparability with mainstream NAR pipelines, we train two acoustic targets under the \emph{same} architecture and schedule: (i) an Fbank variant using 100-dimensional log-mel filterbanks at 24\,kHz with hop length 256, decoded by the Vocos vocoder~\cite{vocos}; and (ii) a Mel--VAE latent variant decoded by BigVGAN~\cite{lee2023bigvgan}.

\noindent\textbf{Datasets.}
We train on the Emilia corpus~\cite{he2024emilia} (approximately $\sim$95k hours of English and Chinese) after filtering transcription errors and other anomalies. Zero-shot TTS is evaluated on three benchmarks: Seed-TTS~\cite{seedtts} test-en (1{,}088 English utterances from Common Voice), Seed-TTS test-zh (2{,}020 Chinese utterances from DiDiSpeech), and a 44.1\,kHz test set of 1{,}000 samples constructed from AISHELL-3~\cite{shi2021aishell}, following the Seed-TTS evaluation protocol.

\noindent\textbf{Baselines.}
We compare against representative AR and NAR systems. \emph{AR models:} CosyVoice~\cite{du2024cosyvoice1}, CosyVoice2~\cite{du2024cosyvoice2}, Spark-TTS~\cite{wang2025sparkttsefficientllmbasedtexttospeech}. \emph{NAR models:} MaskGCT~\cite{maskgct}, E2-TTS~\cite{E2-TTS}, F5-TTS~\cite{chen2024f5}, ZipVoice~\cite{zhu2025zipvoice}.

\noindent\textbf{Metrics.}
We adopt a cross-sentence, zero-shot setting with both reproducible, model-based metrics and human perception scores. Intelligibility is measured by WER using ASR backends: Whisper-large-v3~\cite{whisper-v3} for English and Paraformer-zh~\cite{Paraformer} for Chinese. Speaker similarity (SIM-o) is computed as the cosine similarity between WavLM-based ECAPA-TDNN embeddings~\cite{ECAPA-TDNN} extracted from the prompt and synthesized speech. Naturalness is estimated by UTMOS~\cite{saeki2022utmos}. For human evaluation, we report NMOS (naturalness MOS) and QMOS (quality MOS).

\section{Experimental Results}
\label{sec:Experimental Results}

\subsection{Overall Comparison}
Table~\ref{tab:main_results} reports objective and subjective results across datasets. Compared with representative NAR systems (E2-TTS, F5-TTS, ZipVoice, MaskGCT) and AR systems (CosyVoice, CosyVoice2, Spark-TTS), M3-TTS--Fbank attains strong overall performance: on Seed-TTS en, it achieves the lowest WER among NAR models (1.48) and the highest UTMOS (3.88) with SIM-o 0.681; on Seed-TTS zh, it yields the best SIM-o (0.762), the best UTMOS (3.18), and a competitive WER (1.36, second overall). In double-blind listening tests, M3-TTS--Fbank obtains NMOS/QMOS of 3.80/3.99, surpassing strong NAR baselines (e.g., ZipVoice 3.78/3.95) and closely matching or slightly exceeding ground truth (3.78/3.95). M3-TTS--VAE further attains the lowest WER on Seed-TTS en (1.36) and the lowest NAR WER on Seed-TTS zh (1.31), albeit with lower SIM-o and UTMOS (2.80/2.18). We attribute the gains primarily to Joint--DiT, which aligns text and speech in a unified attention space and avoids alignment bias from filler-token padding and uniform upsampling, thereby improving intelligibility and prosody.

\begin{figure}[t]
    \centering
    \includegraphics[width=0.9\linewidth]{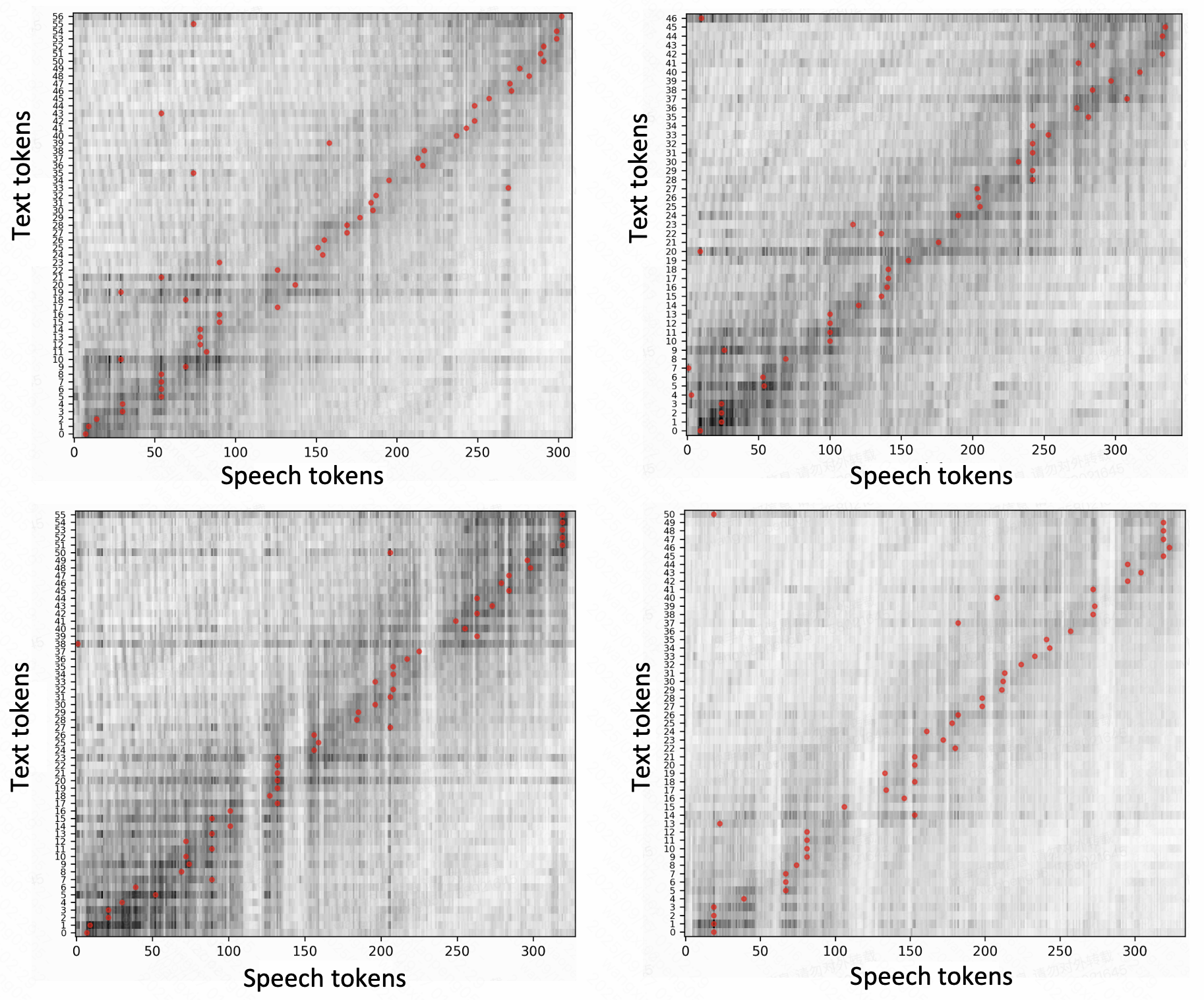}
    \caption{Joint--DiT cross-modal attention visualization for four test samples. Each heatmap shows attention from speech tokens (rows) to text tokens (columns); red dots indicate the row-wise argmax.}
    \label{fig:attn}
\end{figure}

 \begin{table}[t]
    \centering
    \caption{Training time on the Emilia corpus (8$\times$A100, batch size 192), comparing VAE and Fbank features under the MMDiT architecture.}
    \label{tab:efficiency}
    \resizebox{0.45\textwidth}{!}{%
    \begin{tabular}{l c c c}
      \hline
      \textbf{Variant}      & {\textbf{Params} (M)} & {\textbf{Time} (h)} & {\textbf{Speedup} ($\times$)} \\
      \hline
      M3-TTS\textendash Fbank & 355                   & 90                  & 1.0 \\
      M3-TTS\textendash VAE    & 355                   & 31                  & 2.9 \\
      \hline
    \end{tabular}}
    \vspace{-8pt}
\end{table}

To assess the impact of representation choice, we evaluate M3-TTS-VAE on AISHELL3-test (44.1\,kHz) and Seed-TTS (24\,kHz), and include VAE Reconstruction to indicate the codec upper bound. At 44.1\,kHz, M3-TTS-VAE performs below ground truth and the reconstruction ceiling: WER 10.7, SIM-o 0.540, UTMOS 1.78, compared with ground truth and VAE Reconstruction, suggesting constraints from accumulated generation errors and codec capacity under a high sampling rate and cross-corpus evaluation. At 24\,kHz, M3-TTS-VAE often attains lower WER than M3-TTS-Fbank (en 1.36, zh 1.31) but lags on SIM-o/UTMOS. These results indicate that the VAE latent’s roughly $2\times$ temporal compression and distributional prediction ease alignment and regression (benefiting WER), whereas naturalness and timbral detail are limited by codec bandwidth and the reconstruction ceiling; domain shift across corpora and sampling rates further amplifies the gap.


\subsection{Joint Attention Visualization}
Figure~\ref{fig:attn} presents four examples of Joint--DiT cross-modal attention, with speech tokens on the $x$-axis and text tokens on the $y$-axis; brighter intensities indicate larger attention weights, and red dots mark the row-wise argmax (the best speech alignment for each text token). The maps exhibit an approximately monotonic diagonal structure with minimal off-diagonal drift. These qualitative observations suggest that the unified attention space learns stable text–speech alignment.

\subsection{Training Efficiency}
We benchmark training wall-clock time on the Emilia corpus under controlled conditions:
8$\times$A100 GPUs, batch size 192, and an identical training schedule. As shown in
Table~\ref{tab:efficiency}, M3-TTS-Fbank completes in 90\,h,
whereas M3-TTS-VAE completes in 31\,h, yielding a
\textbf{$\sim$3$\times$} speedup. We attribute this gain primarily to the reduced sequence length
and dimensionality in the Mel VAE pathway, which increases throughput without altering
optimization hyperparameters.

\subsection{Discussion}
Despite strong empirical results, M3-TTS has two main limitations. First, Mel-VAE was trained on a small, mixed-sampling-rate corpus, which likely limits latent expressiveness; scaling to larger, better-balanced data with more 44.1 kHz audio should improve fidelity and robustness. Second, inference concatenates the prompt text and reference audio before the target sequence(as in ZipVoice and F5-TTS), which adds latency and limits zero-shot flexibility.

\section{Conclusion}
\label{sec:conclusion}

In this work, we introduced M3-TTS, a multimodal text-to-speech system that combines Mel-VAE latent representations with MMDiT alignment for TTS. Our approach features a Joint-DiT mechanism that enables dynamic cross-modal attention between text and speech, eliminating traditional padding and upsampling limitations. The Mel-VAE codec provides low-dimensional continuous modeling with significant memory reduction. On Seed-TTS and AISHELL-3, M3-TTS attains state-of-the-art non-autoregressive results. In the future, we plan to extend this framework to dialogue.

\vfill\pagebreak

\bibliographystyle{IEEEbib}
{\small
\bibliography{strings,refs}
}

\end{document}